\documentclass[aps,prl,twocolumn,superscriptaddress]{revtex4-1}

\usepackage{amsmath}
\usepackage{graphicx}
\usepackage{epstopdf}
\usepackage{algorithm,algpseudocode}
\usepackage{hyperref}
\usepackage{xcolor}
\begin{document}


\title{A centrality measure for quantifying spread on weighted, directed networks}


\author{Christian G. Fink}
\email[]{finkt@gonzaga.edu}
\affiliation{Dept. of Physics, Gonzaga University}
\author{Kelly Fullin}
\affiliation{Dept. of Computer Science, Ashland University}
\author{Guillermo Gutierrez}
\affiliation{Dept. of Industrial and Systems Engineering, Georgia Institute of Technology}
\author{Nathan Omodt}
\affiliation{Dept. of Mechanical Engineering, Gonzaga University}
\author{Sydney Zinnecker}
\affiliation{Dept. of Physics, Gonzaga University}
\author{Gina Sprint}
\affiliation{Dept. of Computer Science, Gonzaga University}
\author{Sean McCulloch}
\affiliation{Dept. of Computer Science, Ohio Wesleyan University}


\date{\today}

\begin{abstract}
While many centrality measures for complex networks have been proposed, relatively few have been developed specifically for weighted, directed (WD) networks. Here we propose a centrality measure for spread (of information, pathogens, etc.) through WD networks based on the independent cascade model (ICM). While deriving exact results for the ICM requires Monte Carlo simulations, we show that our centrality measure (Viral Centrality) provides excellent approximation to ICM results for networks in which the weighted strength of cycles is not too large. We show this can be quantified with the leading eigenvalue of the weighted adjacency matrix, and we show that Viral Centrality outperforms other common centrality measures in both simulated and empirical WD networks.
\end{abstract}


\maketitle

\section{I. Introduction}

The literature on network centrality measures is vast. Unlike identifying a network's shortest paths \cite{dijkstra1959note} or determining the stability of its synchronization \cite{pecora1998master}, there is no single method for identifying the most influential nodes in a network. This is to be expected, since nodes and connections represent vastly different physical/social entities from one network to the next, and the characteristics which confer centrality often depend on the phenomena of interest \cite{lu2016vital}.

Many centrality measures aim to identify nodes that are influential with  respect to spreading phenomena, such as the spread of infectious disease \cite{newman2002spread}, rumors \cite{nekovee2007theory}, or information \cite{kempe2003maximizing}. The performance of such centrality measures is often determined by comparison with simulations of spread that are highly simplified yet still computationally expensive. The independent cascade and linear threshold models are perhaps the two most popular models for simulating information diffusion through social networks \cite{shakarian2015independent}.

In this paper we propose a centrality measure based on the independent cascade model (ICM), a generalization of the susceptible-infected-recovered (SIR) epidemiological model originally proposed by Goldenberg et al. \cite{goldenberg2001talk,goldenberg2001using}. Like the SIR model, the ICM is used to simulate spread through a network in which nodes may be in one of three states: susceptible, active (or infected), and inactive (or recovered). The ICM runs in discrete time steps, with an active node having just one opportunity to activate its neighbors before becoming inactive. The ICM generalizes the SIR model by specifying network structure using a weighted, directed (WD) adjacency matrix $\mathbf{\mathcal{P}}$, with $\mathcal{P}_{ij}$ giving the probability of node $j$ activating node $i$, given that node $j$ is itself active.

A tremendous amount of work has gone into identifying the optimal set of seed nodes of a given size to maximize the total number of nodes subsequently activated in the ICM \cite{leskovec2007cost, chen2009efficient,chen2010scalable, wang2010community,min2018identifying, altarelli2014containing, morone2015influence}. Kempe et al. showed this to be an NP-Hard problem \cite{kempe2003maximizing}, necessitating heuristic approaches. Such heuristic approaches to this combinatorial optimization problem are not the focus of this study. Instead, we address the simpler problem of ranking individual nodes according to the expected number of others they will infect, given a single activated node as the starting point.

Various centrality measures have been proposed to identify such influential spreaders in the SIR model with homogeneous transmission probability, including degree, k-core, betweenness centrality, eigenvector centrality, and PageRank, among others \cite{lu2016vital}. Such classic approaches suffer from two major drawbacks, however: they do not generally take into account specific details of the spreading model, and they do not apply to WD networks. In cases where a classic centrality measure may be generalized to WD networks, there are often multiple ways of doing so, with no consensus as to which way is best \cite{garas2012k,wei2015weighted,al2017identification}. In general, there are relatively few centrality measures that have been developed specifically for WD networks.

In this study, we propose a novel centrality measure for WD networks specifically designed to identify influential spreaders in the ICM. We call our measure the Viral Centrality (VC), and we show that it gives results nearly to identical to the ICM but much faster, so long as the strength of cycles within a network is small. We show that the strength of cycles within a network may be quantified using the leading eigenvalue of the weighted adjacency matrix, and we show that VC's performance compares favorably to several other WD centrality measures in synthetic and empirical networks. 

\onecolumngrid

\begin{algorithm}[H] 
\caption{Viral Centrality}
\label{alg:vc}
\begin{algorithmic}[1]
\For{each node as SeedNode}
    \State{prob\_susceptible = ones(TotalNodes)} \textcolor{blue}{\# initializing probabilities a given node is still susceptible}
    \State{prev\_activated = zeros(TotalNodes)} \textcolor{blue}{\# initializing probabilities a given node was activated on previous time step}
    \State{cur\_activated = zeros(TotalNodes)} \textcolor{blue}{\# initializing probabilities a given node is activated on current time step}
    \item[]
    \State{prev\_activated[SeedNode] = 1} \textcolor{blue}{\# set seed node to definitely being activated on previous time step...}
    \State{cur\_activated[SeedNode] = 0}  \textcolor{blue}{\# ... and therefore definitely NOT activated on current time step}
    \item[]
    \State{t=1} \textcolor{blue}{\#initialize time step} 
    \While{(termination condition)}
    \item[]
        \State{perform breadth-first search to determine all nodes within t edges of SeedNode}
        \item[]
        \For{each Node within reach of SeedNode}
            \State prob\_uninfected = 1 \textcolor{blue}{\# initialize probability that a given node is \emph{not} infected on this time step}
            \For{each Neighbor sending a connection to Node}
                \State{\textcolor{blue}{\# $\mathcal{P}$[i, j] is probability of node j activating node i, given that node j was activated on previous time step}}
                \State prob\_uninfected = prob\_uninfected * (1-prev\_activated[Neighbor]*$\mathcal{P}$[Node, Neighbor])
            \EndFor
            \State cur\_activated[Node]=(1-prob\_uninfected)*prob\_susceptible[Node]
        \EndFor
        \item[]
        \For{each Node in TotalNodes} \textcolor{blue}{\# clean-up in preparation for next time step}
            \State prev\_activated[Node] = cur\_activated[Node] 
            \State prob\_susceptible[Node] = prob\_susceptible[Node] - cur\_activated[Node]
        \EndFor
        \item[]
        \State t = t + 1
    \EndWhile
   \State viral\_centrality[SeedNode] = sum(1-prob\_susceptible) - 1  \textcolor{blue}{\#``-1'' discounts initial activation of seed node} 
\EndFor
\end{algorithmic}
\end{algorithm}

\twocolumngrid

\section{II. Results}

\subsection{Viral Centrality algorithm}
The VC algorithm aims to calculate the expected total number of nodes activated by a given set of initial seed nodes in the ICM. (In this study, we explore VC as a centrality measure, so we focus on the special case of just one initially activated seed node.) As shown in Algorithm \ref{alg:vc}, for each seed node VC loops over several discrete time steps (Lines 8-22), similar to a Monte Carlo simulation. Rather than prescribe precisely defined states to each node (susceptible, active, or inactive), however, VC continually updates the probability that each nodes remains susceptible (in the array \texttt{prob\_susceptible}). At the end of the algorithm (Line 23), taking one minus this value gives the probability a given node was activated at some previous point in time, and adding all these probabilities gives the expected number of nodes ultimately activated by the seed node. 

Note how Line 8 allows one to specify a condition for terminating the algorithm. This could be simply a fixed number of time steps, or a condition based on the convergence of \texttt{prob\_susceptible} to stable values. We later explore the consequences of both approaches in applying VC to empirical networks.

The VC algorithm gives exact results for acyclic networks, but strictly overestimates the expected number of nodes infected for networks with cycles. This can be seen most easily with a simple example: suppose node $i$ and node $j$ are reciprocally connected, $i \longleftrightarrow j$, and further suppose node $j$ has no connections with any other nodes. If node $i$ activates node $j$, the ICM does not allow $j$ to subsequently activate node $i$ (because node $i$ is activated for only one time step, and inactivated thereafter). Yet the VC algorithm allows node $j$ to increase node $i$'s probability of being activated, resulting in an overestimate of node $i$'s expected activation probability. While this simple example is for a 2-cycle, it applies to arbitrary $n$-cycles. It has been shown that properly accounting for all such cycles results in an algorithm that scales exponentially with the number of nodes, making it impractical for large networks \cite{yang2019influence}.

\subsection{Condition for accuracy}
Fortunately, it is possible to determine the credibility of VC's results for a given network by considering the leading eigenvalue of the network's weighted adjacency matrix, $\mathbf{\mathcal{P}}$. To see this, we first devise a measure to quantify the ``strength of cycles'' within a WD network. First, the number of paths of length $\ell$ from node $j$ to node $i$  in an unweighted network with adjacency matrix $\mathbf{A}$ is equal to $(\mathbf{A}^{\ell})_{ij}$, and therefore the number of cycles of length $\ell$ starting and ending on node $i$ is $(\mathbf{A}^{\ell})_{ii}$. For a WD network, we may replace the unweighted adjacency matrix $\mathbf{A}$ with the weighted adjacency matrix $\mathbf{\mathcal{P}}$, understanding that each cycle is then weighted by the product of its links. 

A reasonable measure of the strength of cycles in a network, $SC$, is then found by summing over all nodes and all possible cycle lengths:

\begin{equation}
    SC = \sum_{\ell=2}^{\infty} \mathrm{Tr} (\mathbf{\mathcal{P}}^{\ell})
\label{eq:cycles1}
\end{equation}

We may then use two properties from linear algebra to further simplify this result. First, the trace of a matrix is equal to the sum of its eigenvalues. Second, if $\lambda$ is an eigenvalue of a matrix $\mathbf{M}$, then $\lambda^{n}$ is the corresponding eigenvalue of $\mathbf{M}^{n}$. 

For an $n$-node network, Eq. \ref{eq:cycles1} then becomes

\begin{equation}
    SC = \sum_{\ell=2}^{\infty} \lambda_1^{\ell} + \lambda_2^{\ell} + \dots + \lambda_n^{\ell}.
\label{eq:cycles2}
\end{equation}

This measure will diverge when the magnitude of the weighted adjacency matrix's leading eigenvalue is greater than or equal to 1. In this case, the error (relative to ground truth ICM results) in VC's predicted number of nodes activated will be hopelessly large. On the other hand, VC gives results exactly equal to ICM results in acyclic networks, for which $SC=0$. We expect a smooth transition in VC's performance between these two extremes, and predict low error between VC and ICM results when $|\lambda|_{\mathrm{max}} \ll 1$.

\subsection{Results for Erd\H{o}s-R\'{e}nyi networks}

We explored VC's accuracy as a function of $|\lambda|_{\mathrm{max}}$ in a simple Erd\H{o}s-R\'{e}nyi model. We synthesized an assortment of WD 1000-node networks by specifying two different probability values: $p_{\mathrm{con}}$ was the probability of a potential directed connection actually being instantiated, and $p_{\mathrm{trans}}$ was the transmission probability (fixed to the same constant value for all connections within any particular network). We independently swept over a range of 0.005 to 0.038 for both probabilities (this range was chosen in order to give values for $|\lambda|_{\mathrm{max}}$ ranging from roughly 0 to 1). For each combination of $p_{\mathrm{con}}$ and $p_{\mathrm{trans}}$, we generated a corresponding network, then ran ICM Monte Carlo simulations and compared these ``ground truth'' results to those predicted by VC. We then computed the mean relative error between the ICM and VC results, defined as 

\begin{equation}
    \epsilon = \frac{\sum_{i=1}^{N} \frac{\lvert a^{\mathrm{ICM}}_i - a^{\mathrm{VC}}_i \rvert}{a^{\mathrm{ICM}}_i}}{N},
\end{equation}

\noindent
where $a^{\mathrm{ICM}}_i$ is the mean number of nodes activated by node $i$ in the ICM (averaged across 10,000 Monte Carlo trials) and $a^{\mathrm{VC}}_i$ is the expected number of nodes activated by node $i$ according to Viral Centrality. For Erd\H{o}s-R\'{e}nyi networks, the VC algorithm was terminated when either of the following conditions was first met: 1) the maximum relative error (from one time step to the next for a particular node, with the maximum taken across all nodes) first dipped below $10^{-5}$, or 2) twenty time steps were completed.

\begin{figure*}
\includegraphics{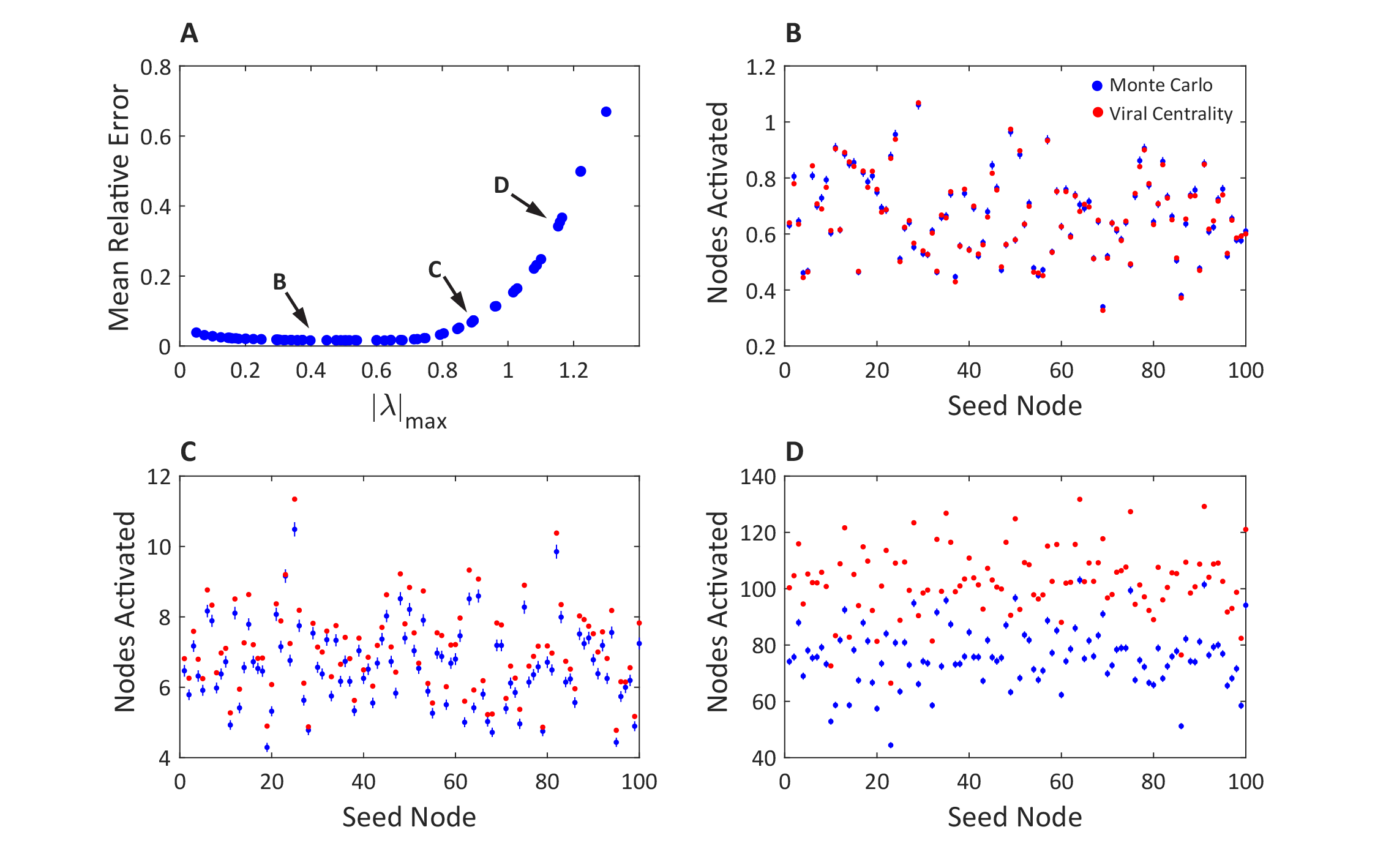}
\caption{Performance of Viral Centrality in Erd\H{o}s-R\'{e}nyi networks. Various combinations of connection probability ($p_{\mathrm{con}}$) and transmission probability ($p_{\mathrm{trans}}$ ) were used to synthesize 1000-node networks. 10,000 Monte Carlo simulations of the Independent Cascade Model were conducted for each network, and the average number of nodes activated for each seed node was compared to that predicted by Viral Centrality. A: Each point represents a single network, with different points implying different combinations of $p_{\mathrm{con}}$ and $p_{\mathrm{trans}}$. The mean relative error between ICM and VC is plotted as a function of the magnitude of the leading eigenvector of each network's adjacency matrix. B: Results for a network with $p_{\mathrm{con}}=0.02$ and $p_{\mathrm{trans}}=0.02$. The figure shows the mean (ICM Monte Carlo) or expected (VC) number of nodes activated by 100 randomly selected seed nodes. Error bars for Monte Carlo results represent standard error of the mean over 10,000 trials. C: Same as B, but for a network with $p_{\mathrm{con}}=0.03$ and $p_{\mathrm{trans}}=0.03$. D: Same as B and C, but for a network with $p_{\mathrm{con}}=0.034$ and $p_{\mathrm{trans}}=0.034$.}
\label{fig:erdos_renyi}
\end{figure*}

Fig. \ref{fig:erdos_renyi}A shows how the mean relative error depended on the magnitude of the leading eigenvalue of the weighted adjacency matrix. As expected, VC performed very well for $|\lambda|_{\mathrm{max}} \ll 1$, and the error steadily increased as $|\lambda|_{\mathrm{max}}$ approached and exceeded 1. (Note that the error did not go to zero as $|\lambda|_{\mathrm{max}}$ approached zero due to imprecision in the Monte Carlo results, resulting from a finite number of ICM trials for each network.) Panels B, C, and D compare Monte Carlo and VC results for 100 randomly selected nodes (out of 1000) in three representative networks. (We do not show results for all 1000 seed nodes in order to aid visualization.) Note the excellent match in B (small $|\lambda|_{\mathrm{max}}$) and the worsening performance as $|\lambda|_{\mathrm{max}}$ increases (C and D). Also note how VC tends to overestimate the number of nodes activated.

\subsection{Congressional Twitter network}

We applied VC to an empirical network by using Twitter's API to construct the Twitter interaction network for the 117th United States Congress House of Representatives. We quantified empirical transmission probabilities according to the fraction of times one member retweeted, quote tweeted, replied to, or mentioned another member's tweet. (See Methods section for more details.) Requiring a minimum of 100 Tweets over a four-month span gave a network of 475 Congressional members, with transmission probabilities ranging between 0.00053 and 0.13, and a leading eigenvalue for the adjacency matrix of 0.208.

\begin{figure*}
\includegraphics{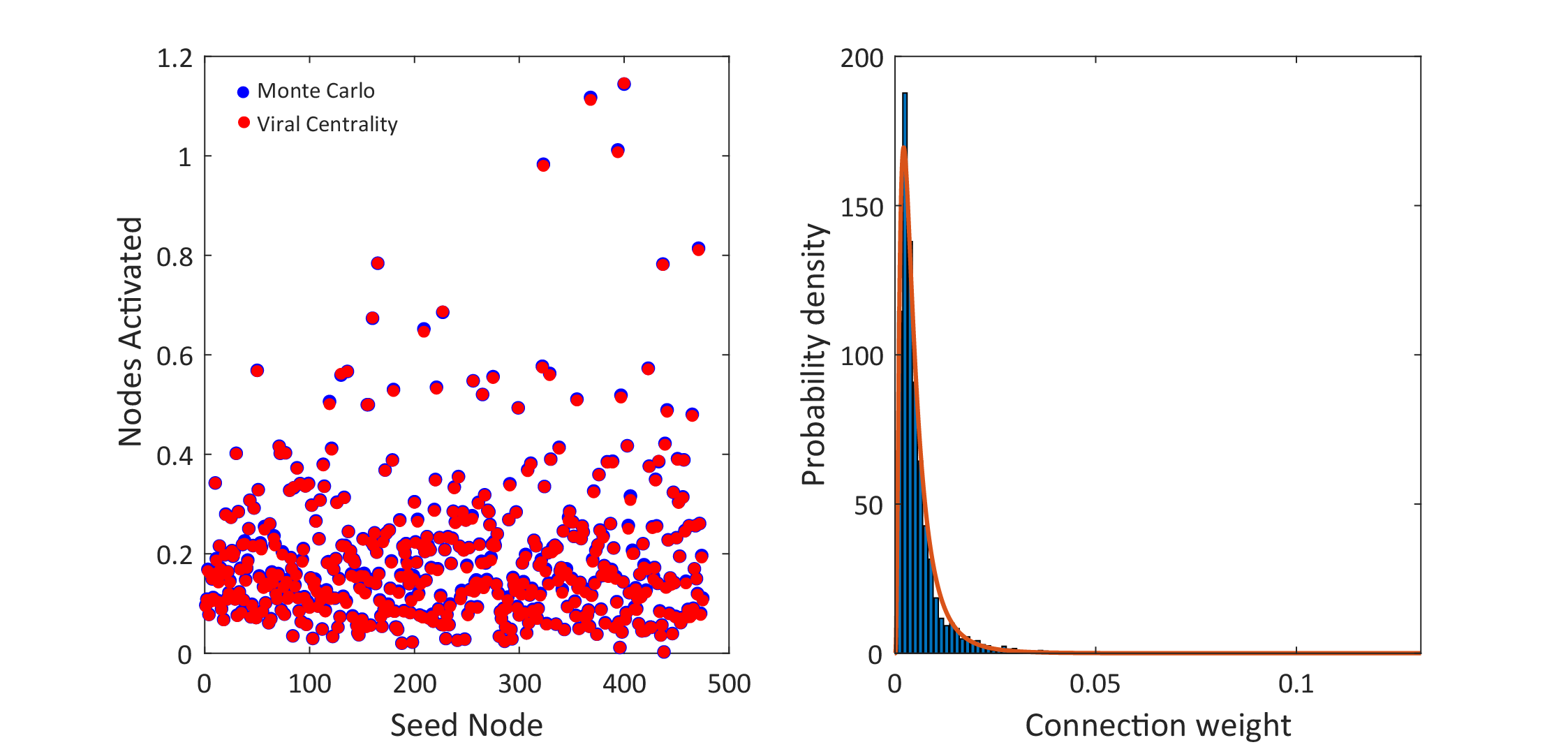}
\caption{Viral Centrality and Independent Cascade Model results for Congressional Twitter network. A: Mean (ICM Monte Carlo) or expected (VC) number of nodes activated by each of the 475 seed nodes. Monte Carlo results represent averages over $10^6$ trials, with error bars displayed but too small to be seen. B: Distribution of transmission probabilities for this network. Red line represents fit by a lognormal distribution.}
\label{fig:congress_twitter}
\end{figure*}

Fig. \ref{fig:congress_twitter}A shows excellent agreement between VC and Monte Carlo ICM results. (For this figure, the VC computation terminated with a tolerance of 0.001, meaning every seed node's expected number of nodes activated had to change by less than 0.1\% from the previous iteration.) Fig. \ref{fig:congress_twitter}B shows the distribution of transmission probabilities (i.e., connection weights) was well approximated by a lognormal distribution (see Methods section), which we used later in formulating the Higgs Twitter dataset (discussed in the next section). 


In Table \ref{tab:congress_table}, we compare the performance of VC to other common centrality measures in ranking the influence of members of Congress on Twitter. We use expected number of nodes activated in Monte Carlo ICM simulations for the ground truth ranking, then compute Kendall's tau correlation \cite{kendall1938new} with the rankings obtained by WD versions of degree, k-core \cite{garas2012k,wei2015weighted,al2017identification}, and PageRank \cite{zhang2022pagerank}. We also compare to diffusion degree, which is a generalization of weighted degree that takes second-order neighbors into account \cite{kundu2011new,pal2014centrality,gaye2015new}. 

The table shows how VC outperforms all these centrality measures, albeit at the cost of taking much longer to compute. Diffusion degree comes closest to matching the performance of VC, giving a Kendall's tau value nearly identical to VC when ranking all nodes, but clearly inferior when ranking just the top 10\% of nodes. It is perhaps not surprising that diffusion degree's performance is closest to VC, since it is similar (but not identical) to running VC for two time steps. Diffusion degree starts with the weighted out-degree of a node (first-order neighbors), then adds the products of the two connection weights required to reach each second-order neighbor. This is similar to two-step VC, except that diffusion degree's simplistic approach does not attempt to make the second-order contributions equal to probabilities of activation of second-order neighbors (although for small transmission probabilities, it gives a good approximation). We therefore expect that if we run VC for two time steps (rather than to a convergence tolerance), it will give results roughly on par with those of diffusion degree, and that running for more time steps will further improve performance. Table \ref{tab:congress_table} confirms this expectation, demonstrating that running VC for a small, fixed number of time steps can expedite the VC algorithm at the cost of diminished accuracy.

\begin{table*}[htbp]
  \centering
  \begin{tabular}{|c|c|c|c|c|}
    \hline
    Measure & \begin{tabular}[c]{@{}c@{}}Kendall's Tau \\ (top 10\%)\end{tabular} & \begin{tabular}[c]{@{}c@{}}Kendall's Tau\\(all nodes)\end{tabular} & \begin{tabular}[c]{@{}c@{}} Mean Rel. \\ Error \end{tabular} &\begin{tabular}[c]{@{}c@{}}Compute \\ time (s) \end{tabular} \\
    \hline
    WD Out-degree & 0.882 & 0.949 & N/A & $<10^{-7}$ \\ \hline
    WD k-core & 0.223 & 0.672 & N/A &0.031 \\ \hline
    WD PageRank & 0.415 & 0.515 & N/A &0.047 \\ \hline
    Diffusion degree & 0.882 & 0.989 &  0.025 &0.125 \\ \hline
    \begin{tabular}[c]{@{}c@{}}Viral Centrality \\ (tol=0.001)\end{tabular} & 0.994 & 0.991 & 0.017 &5.03\\ \hline
    \begin{tabular}[c]{@{}c@{}}Viral Centrality \\ (t=2)\end{tabular} & 0.965 & 0.990 & 0.029 & 2.89\\ \hline
    \begin{tabular}[c]{@{}c@{}}Viral Centrality \\ (t=3)\end{tabular} & 0.994 & 0.997 & 0.0047 &7.58\\ \hline
\end{tabular}
\caption{Comparison of VC to other centrality measures for the Congressional Twitter network. Kendall's tau correlation between Monte Carlo ($10^6$ trials per seed node) and centrality rankings was computed for all nodes in the network, as well as for just the top 10\% of nodes in the network. Mean relative error was also computed between average number of nodes activated in Monte Carlo simulations and the number predicted by diffusion degree and VC. VC outperformed all other measures, though at the cost of substantially increased runtime. Monte Carlo simulations required approximately 6 compute-hours. All compute times were determined using the Neuroscience Gateway computing cluster.}
\label{tab:congress_table}
\end{table*}



\subsection{Higgs networks}

\begin{figure}
\includegraphics{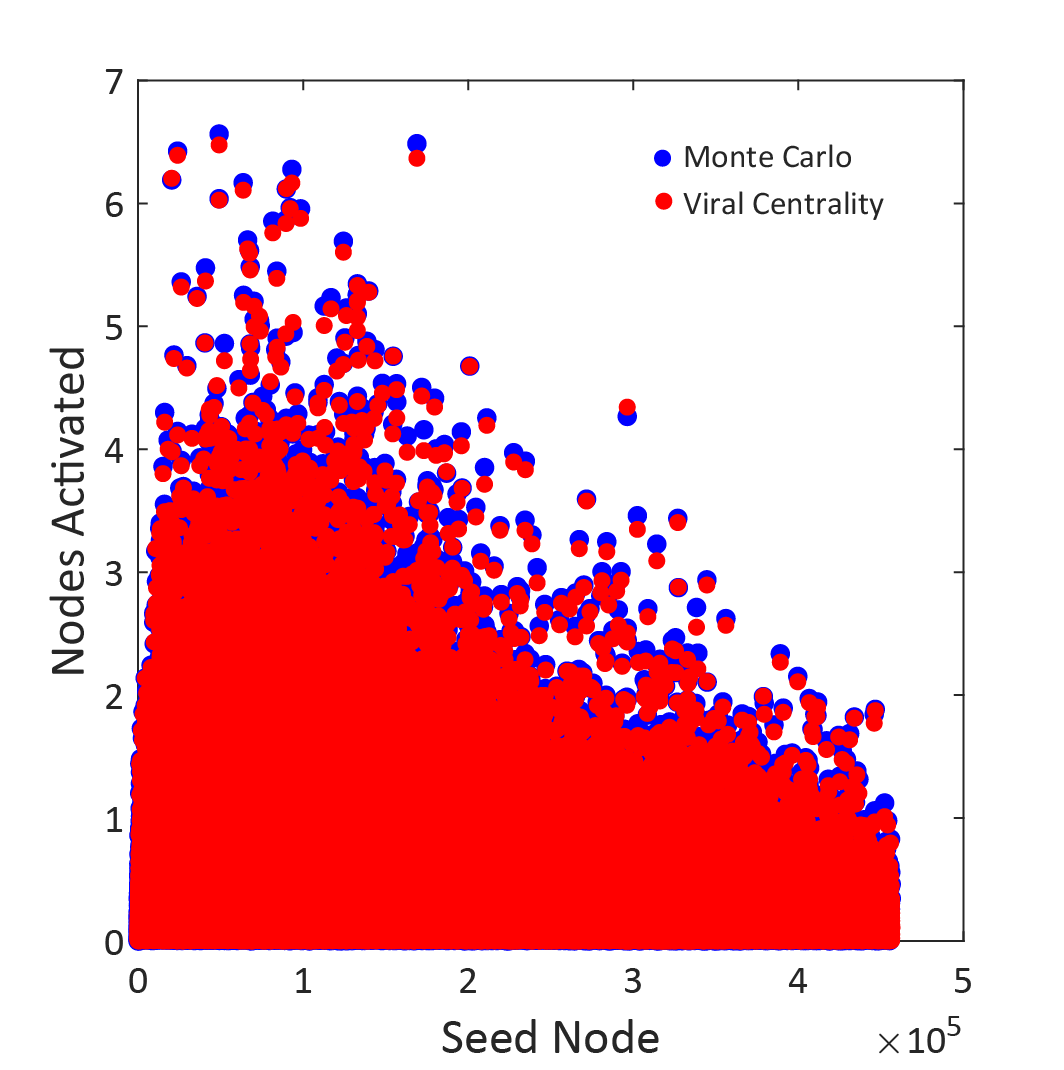}
\caption{Viral Centrality and Independent Cascade Model results for Higgs Twitter network. Plot displays mean (ICM Monte Carlo) or expected (VC) number of nodes activated by each of the 456,626 seed nodes. Monte Carlo results represent averages over $10^4$ trials, with error bars displayed but too small to be seen.}
\label{fig:higgs_twitter}
\end{figure}

We applied VC to the Higgs Twitter dataset, a much larger network of 456,626 nodes derived from Tweets related to the discovery of the Higgs boson in 2012 \cite{de2013anatomy} (see Methods for more details). Fig. \ref{fig:higgs_twitter} overlays the average number of nodes activated according to Monte Carlo ICM and the expected number according to VC (run with a convergence tolerance of 0.001). Table \ref{tab:higgs_table} shows the Kendall's tau between these two results was 0.967 for all nodes, and 0.927 for the top 10\% of nodes. Similar to the Congressional network, the diffusion degree matched this performance for all nodes but was inferior for the top 10\%. VC run with two time steps gave virtually identical results to diffusion degree, whereas running it with three time steps gave even better performance than with the convergence tolerance (suggesting that most seed nodes required only two time steps to converge). 

While VC is obviously much more computationally expensive than more traditional centrality measures (requiring about 400 compute-hours when run with convergence tolerance), it still provided roughly an order of magnitude speed-up over Monte Carlo simulations, which required approximately 5500 compute-hours to complete 10,000 trials per seed node.


\begin{table*}[htbp]
  \centering
  \begin{tabular}{|c|c|c|c|c|}
    \hline
    Measure & \begin{tabular}[c]{@{}c@{}}Kendall's Tau\\(top 10\%)\end{tabular}   & \begin{tabular}[c]{@{}c@{}}Kendall's Tau\\all nodes)\end{tabular} & \begin{tabular}[c]{@{}c@{}} Mean Rel. \\ Error \end{tabular} & \begin{tabular}[c]{@{}c@{}} Compute \\ time (hr) \end{tabular} \\
    \hline
    WD Out-degree & 0.872 & 0.702 & N/A & 5.0 $\times 10^{-5}$ \\ \hline
    WD k-core & 0.355 & -0.017 & N/A & 2.40 \\ \hline
    WD PageRank & 0.316 & 0.144 & N/A & 26.1 \\ \hline
    Diffusion Degree & 0.877 & 0.967 & 0.113 & 0.08 \\ \hline
    Viral Centrality (tol=0.001) & 0.927 & 0.967 & 0.093 & 398.9 \\ \hline
    Viral Centrality (t=2) & 0.877 & 0.967 & 0.113 & 202.5 \\ \hline
    Viral Centrality (t=3) & 0.938 & 0.977 & 0.058 & 789.0 \\ \hline
\end{tabular}
\caption{Comparison of VC to other centrality measures for the Higgs Twitter network. Kendall's tau correlation between Monte Carlo (10,000 trials per seed node) and centrality rankings was computed for all nodes in the network, as well as for just the top 10\% of nodes in the network. Mean relative error was also computed between average number of nodes activated in Monte Carlo simulations and the number predicted by diffusion degree and VC. As in the Congressional network, VC generally outperformed all other measures, though at the cost of substantially increased runtime. Monte Carlo simulations required approximately 5500 compute-hours.}
\label{tab:higgs_table}
\end{table*}

\section{III. Discussion}
In this study we have proposed a novel centrality measure, Viral Centrality (VC), for weighted, directed (WD) networks. VC is designed to identify influential spreaders within networks for which probability of transmission between nodes can be quantified. We have shown that VC well approximates the expected number of nodes activated in the Independent Cascade Model (ICM), so long as the leading eigenvalue of the WD adjacency matrix is much less than 1. Because VC is modeled after the ICM, it is more accurate than other traditional centrality measures in ranking nodes with respect to the number activated in Monte Carlo ICM simulations.

The trade-off for this improved accuracy is slower run time. VC is much slower than degree-based centrality measures, and is too slow to be used as a heuristic to find an optimal set of multiple seed nodes in the ICM \cite{sumith2018influence,leskovec2007cost}.  For such a combinatorial optimization problem, faster centrality measures such as diffusion degree \cite{pal2014centrality} are required for large-scale networks. VC is best applied to networks composed of up to tens of thousands of nodes.

Nevertheless, we have shown that VC is generally more accurate than diffusion degree,
making it an ideal choice for situations where accuracy is paramount, yet speed much faster than Monte Carlo simulations is still required. We have also shown that this superior accuracy can be expected when the leading eigenvalue of a network's WD adjacency matrix is much less than 1 (Fig. \ref{fig:erdos_renyi}A suggests $|\lambda|_{\mathrm{max}} = 0.5$ is a conservative and safe cut-off). Although the calculation of the leading eigenvalue of a matrix is generally expensive, we note that Gershgorin's Circle Theorem may be used to place an upper bound on the leading eigenvalue of an adjacency matrix based purely on its row sums. This provides an inexpensive, sufficient condition for trusting the accuracy of VC's results. Our results for the Congressional Twitter network suggest that transmission probabilities in empirical networks may often be small enough for VC to well approximate ICM results.


It should be noted that there are many generalizations and extensions of the ICM that our centrality measure does not address \cite{kempe2003maximizing,jung2012irie,qin2017efficient,liu2012time}. Further research is necessary to develop efficient algorithms for identifying influential spreaders in these more complex models.


\section{IV. Methods}

\subsection{Comparison centrality measures}

\subsubsection{WD PageRank}
We used an extension of PageRank to WD networks proposed by Zhang et. al. \cite{zhang2022pagerank}. The original PageRank (PR) algorithm \cite{brin1998anatomy} applied to unweighted networks, and defined PR iteratively according to 

\begin{equation}
PR(i) = \alpha \sum_j \frac{a_{ij}}{d^{\mathrm{out}}_j} PR(i) + \frac{1-\alpha}{n},
\end{equation}

\noindent
where $a_{ij}$ is a binary value (0 or 1) specifying whether or not a connection exists from $j$ to $i$. The extension to WD networks is achieved by introducing a second parameter, $\theta \in [0,1]$, that determines the degree to which the centrality score is determined by the weighted adjacency matrix ($w_{ij}$) versus the unweighted one ($a_{ij}$). Calling the WD PR centrality score $\phi$, it may be computed iteratively according to:

\begin{equation}
\phi(i) = \alpha \sum_j \left( \theta \frac{w_{ij}}{s^{\mathrm{out}}_j} + (1-\theta) \frac{a_{ij}}{d^{\mathrm{out}}_j} \right) \phi(i) + \frac{1-\alpha}{n},
\end{equation}

\noindent
where $s^{\mathrm{out}}_j$ is the weighted out-degree of node $j$. Note that in this approach, nodes ``score points'' according to the \emph{incoming} connections they receive. However, in this study influence was based on outgoing connections, so we simply ran the WD PR algorithm on the transpose of our adjacency matrices. Also, in this study we set $\theta=0.5$.

\subsubsection{WD k-core}
Inspired by \cite{garas2012k,wei2015weighted,al2017identification}, we implemented a WD version of k-core decomposition using the following approach:

\begin{itemize}
  \item Calculate the weighted out-degree, $s^{\mathrm{out}}$, of each node in the network.
  \item Determine the smallest weighted out-degree, $s^{\mathrm{out}}_{\mathrm{min}}$.
  \item Remove all nodes for which $s^{\mathrm{out}}=s^{\mathrm{out}}_{\mathrm{min}}$.
  \item Update the values of $s^{\mathrm{out}}$ for the nodes that remain in the network.
\end{itemize}

\noindent
This process terminates when all nodes in the network have been removed. Nodes are ranked according to the order in which they are removed from the network, with those surviving the longest ranked the highest.

\subsubsection{Diffusion degree}
The diffusion degree, DD, generalizes weighted degree to include second-order neighbors in addition to first-order neighbors. Given a weighted adjacency matrix $w_{ij}$, it is defined mathematically by

\begin{equation}
DD(i) = \sum_j w_{ji} \times \left(1 + \sum_k w_{kj} \right).
\end{equation}

\subsection{Datasets}

\subsubsection{Congressional Twitter network}

The Twitter interaction network for the 117th United States Congress House of Representatives was constructed by first obtaining members' official Twitter handles from \url{https://pressgallery.house.gov/member-data/members-official-twitter-handles}. The Twitter API was then used to obtain all Tweets by members of Congress between February 9, 2022, and June 9, 2022. (The latter date corresponds to when the data were requested, and the earlier date was the farthest point backward in time at which one user's number of Tweets exceeded the 3200-Tweet limit set by the API.) Only the 475 representatives (out of 535) who issued 100 or more tweets during this time frame were included in the network.

A weighted, directed network was constructed such that the connection from user $j$ to user $i$ was given by

\begin{eqnarray}
\mathcal{P}_{ij} = \frac{n^{\mathrm{retweet}}_{ij} + n^{\mathrm{quote}}_{ij} + n^{\mathrm{reply}}_{ij} + n^{\mathrm{mention}}_{ij} }{N^{\mathrm{tweets}}_j},
\end{eqnarray}
\noindent
where $n^{\mathrm{retweet}}_{ij}$ is the number of times user $i$ retweeted user $j$, $n^{\mathrm{quote}}_{ij}$ is the number of times user $i$ quote tweeted user $j$, $n^{\mathrm{reply}}_{ij}$ is the number of times user $i$ replied to user $j$, $n^{\mathrm{mention}}_{ij}$ is the number of times user $i$ mentioned user $j$, and $N^{\mathrm{tweets}}_j$ is the total number of Tweets issued by user $j$. The weighted connection, $\mathcal{P}_{ij}$, can therefore be interpreted as the probability that user $i$ is observably influenced by user $j$.

The distribution of non-zero elements $w$ of $\mathcal{P}_{ij}$ was approximated by a lognormal distribution with pdf 

\begin{equation}
p(w) = \frac{1}{sw \sqrt{2\pi}} \; \mathrm{exp}\left( -\frac{\mathrm{log}^2(w)}{2s^2}\right).
\end{equation}

\noindent
The parameter $s=0.7893$ was found to best fit to data (see Fig. \ref{fig:congress_twitter}B).

\subsubsection{Higgs Twitter dataset}
The Higgs Twitter dataset is derived from all tweets from July 1 to July 7, 2012, that contained the following keywords or hashtags: \texttt{lhc}, \texttt{cern}, \texttt{boson}, \texttt{higgs} \cite{de2013anatomy}. The dataset is available for download on the Stanford Network Analysis Project website (\url{https://snap.stanford.edu/data/higgs-twitter.html}). While the original dataset includes information about retweets, replies, and mentions, we chose not to use this information in constructing our network because most users issued only a few tweets in total, resulting in poor statistics for inferring transmission probabilities. 

Instead, we started with the follower network, which consisted of 456,626 nodes and 14,855,842 directed connections. (If user $j$ followed user $i$, this constituted a connection from node $i$ to node $j$, consistent with the presumed direction of influence.) We then randomly assigned weights (i.e., transmission probabilities) to each connection, using the lognormal distribution empirically determined from the Congressional Twitter network (see Fig. \ref{fig:congress_twitter}B). The magnitude of the largest eigenvalue of the resulting adjacency matrix was 0.074.

\subsection{Acknowledgements}
This work was supported by the Gonzaga Science Research Program, the Ohio Wesleyan Summer Science Research Program, and the National Science
Foundation (1658998). Simulations were conducted on the Neuroscience Gateway computing cluster.


\bibliographystyle{apsrev4-2}
\bibliography{VC_refs}

\end{document}